\documentclass[aps,prb,superscriptaddress,citeautoscript,twocolumn,notitlepage]{revtex4-1}

\usepackage{graphicx}
\usepackage{amsmath}
\usepackage{amssymb}
\usepackage{braket}
\usepackage[usenames]{color}

\newcommand{\bea}{\begin{eqnarray*}}
\newcommand{\eea}{\end{eqnarray*}}
\newcommand{\bne}{\begin{equation*}}
\newcommand{\ede}{\end{equation*}}

\newcommand{\bnen}{\begin{equation}}
\newcommand{\eden}{\end{equation}}
\newcommand{\bean}{\begin{eqnarray}}
\newcommand{\eean}{\end{eqnarray}}
\newcommand{\bnsn}{\begin{subequations}}
\newcommand{\edsn}{\end{subequations}}

\newcommand{\bna}{\begin{array}}
\newcommand{\eda}{\end{array}}
\newcommand{\bnm}{\begin{enumerate}}
\newcommand{\edm}{\end{enumerate}}

\newcommand{\ttmatrix}[4]{\left(\bna{cc} #1 & #2 \\ #3 & #4 \eda\right)}

\renewcommand{\vec}[1]{\text{\boldmath{$ #1 $}}}

\begin{document}
\title{Poor man's topological quantum gate based on the 
Su-Schrieffer-Heeger model}

\author{P\'eter Boross}
\email{boross.peter@wigner.mta.hu}
\affiliation{Institute for Solid State Physics and Optics, Wigner Research Centre for Physics,
Hungarian Academy of Sciences, H-1525 Budapest P.O. Box 49, Hungary}

\author{J\'anos K. Asb\'oth}
%\email{palyi@mail.bme.hu}
\affiliation{Institute for Solid State Physics and Optics, Wigner Research Centre for Physics,
Hungarian Academy of Sciences, H-1525 Budapest P.O. Box 49, Hungary}

\author{G\'abor Sz\'echenyi}
%\email{palyi@mail.bme.hu}
\affiliation{Department of Materials Physics, E\"otv\"os Lor\'and University, 
P\'azm\'any P\'eter s\'et\'any 1/A, H-1117 Budapest, Hungary}

\author{L\'aszl\'o Oroszl\'any}
\affiliation{Department of Physics of Complex Systems, 
E\"otv\"os Lor\'and University, 
P\'azm\'any P\'eter s\'et\'any 1/A, H-1117 Budapest, Hungary}
\affiliation{MTA-BME Lend\"ulet Topology and Correlation Research Group,
Budapest University of Technology and Economics, 
Budafoki \'ut 8., H-1111 Budapest, Hungary}

\author{Andr\'as P\'alyi}
\email{palyi@mail.bme.hu}
\affiliation{
MTA-BME Lend\"ulet Exotic Quantum Phases Research Group
and 
Department of Theoretical Physics, 
Budapest University of Technology and Economics, 
Budafoki \'ut 8., H-1111 Budapest, Hungary}

\date{\today}

\begin{abstract}
Topological properties of quantum systems could provide
protection of information against environmental noise, and thereby 
drastically advance their potential 
in quantum information processing. 
Most proposals for topologically protected quantum gates are based on
many-body systems, e.g., fractional quantum Hall states, 
exotic superconductors, or ensembles of interacting spins, 
bearing an inherent conceptual complexity. 
Here, we propose and study a topologically protected quantum
gate, based on a one-dimensional
single-particle tight-binding model, known as 
the Su-Schrieffer-Heeger chain. 
The proposed $Y$ gate acts
in the two-dimensional 
zero-energy subspace of a Y junction 
assembled from three chains,
and is based on the spatial exchange of the defects 
supporting the zero-energy modes.
With numerical simulations, we demonstrate
that the gate is robust against hopping disorder but
is corrupted by disorder in the on-site energy. 
Then we show that this robustness is topological 
protection, and that it arises as a joint consequence of 
chiral symmetry, time-reversal symmetry
and the spatial separation of the zero-energy
modes bound to the defects. 
This setup will most likely not lead to a practical 
quantum computer, nevertheless it does provide
valuable insight to aspects of
topological quantum computing as an elementary minimal model.
Since this model is non-interacting and non-superconducting, 
its dynamics can be studied experimentally, e.g., using coupled
optical waveguides.
\end{abstract}

%\pacs{73.63.Fg}
%73.63.Fg Nanotubes 

\maketitle

%\tableofcontents

\section{Introduction}

A generic task in quantum information processing is to
perform quantum gates.
In the theory of topological quantum computing\cite{Nayak_rmp,Pachos}, 
a very general framework is to encode 
and manipulate quantum information
in a quantum system with \emph{defects}.
In this framework, the presence of the defects ensures
the existence of an energy-degenerate eigensubspace
of the Hamiltonian, quantum information is encoded
in this subspace, and gates are performed by 
exchanging
the spatial positions of the
defects in a slow, adiabatic fashion\cite{Ivanov,Alicea}. 
The microscopic details of
the braiding of the defects is not relevant, 
instead it is only the topology of the world lines
of the defects which determines the quantum gate being
performed. 
This implies that quantum information and
quantum gates in these models are protected
from certain types of perturbations, which is often phrased as 
\emph{topological protection}.

In most cases, topological quantum computing 
has been studied via exotic, 
strongly correlated interacting quantum
systems.
One example is the toric code\cite{KitaevToricCode};
in its simplest form, it is 
a two-dimensional lattice of localized 
interacting spin-1/2 particles
with four-spin interactions.
One-dimensional topological superconductors are also expected to
provide a platform for topological quantum computing.
%in that case however, interactions
%are hidden in the mean-field description and the system is 
%described by a `free' Hamiltonian that is quadratic in fermion
%operators, but includes anomalous terms that do
%conserve the number parity of the particles but do not
%conserve the number of particles\cite{KitaevChain}.
In this one-dimensional example, the defects are also called
domain walls, as they separate topological and 
trivial sample segments from each other. 
Quantum gates based on braiding of defects in a
Y-junction based on such superconductors\cite{Alicea} 
(\emph{Majorana Y-junction})
have topological protection: even in the presence of local 
perturbations, 
gate errors are expected to be suppressed
if the speed of the exchange is reduced, 
and the length of the system is increased so that the
defects are more and more separated.
In this example, topological protection is guaranteed
as the joint consequence of the spatial separation 
of the defects and the particle-hole symmetry of the
Bogoliubov-de Gennes Hamiltonian governing the dynamics. 

Mostly because of the expected paradigm shift in quantum computing
research, great efforts are devoted to the experimental 
realization of topological quantum gates in Majorana Y-junctions
based on hybrid superconductor-semiconductor 
nanostructures\cite{Mourik,LutchynReview}.
This is, however, a rather challenging path, 
because of the inherent complexity of the nanofabrication 
procedures and the state-of-the-art low-temperature electronic
measurements required.

Theory works have already mentioned
the potential connection between non-interacting, non-superconducting,
single-particle tight-binding lattice models and topological
quantum computing\cite{Klinovaja,ZhigangSong,Droth}.
One intriguing scheme, that of vortex-like defects in the Kekule distortion of 
a hexagonal lattice, which behave similarly to Majorana zero modes in 
p-wave superconductors\cite{hou2007electron,iadecola2016non}, has even 
been recently realized in optical waveguide 
arrays\cite{menssen2019photonic}. In this work, we introduce and study an 
even simpler model system, requiring only a few sites  
without interactions or superconductivity,  which can 
realize quantum gates that enjoy a level of topological protection. 
%These works, however, left it as an important open question if topologically
%protected quantum gates can be performed in such simple 
%systems. 
%In this work, we introduce and study a model quantum system 
%with topologically protected quantum gates,
%without the need of any interactions or
%anomalous superconducting Hamiltonian terms.
The setup studied here is based on the 
Su-Schrieffer-Heeger (SSH) 
model\cite{BarisicPRB1,SSH,Asboth}, 
in which a single particle
lives in a simple
tight-binding lattice with 
a two-orbital unit cell.

In particular, we propose a setup and a braiding protocol 
which resembles the Majorana Y-junction scheme\cite{Alicea,Sekania}.
Similarly to the latter, in our SSH Y-junction, the exchange of defects 
provides a single-qubit rotation in a two-dimensional degenerate subspace
of the Hamiltonian.
The gate in our SSH Y-junction is a $Y$ gate, corresponding to a rotation
by angle $\pi$, in contrast to the gate 
in the Majorana Y-junction which provides a rotation of angle $\pi/2$. 
We demonstrate that the $Y$ gate in 
the SSH Y-junction is topologically protected, if, 
throughout the duration of the braiding protocol, 
the spatial separation of the defects is large, and the chiral symmetry 
and the time-reversal symmetry of the Hamiltonian are 
maintained.

Note that we do not claim that 
the model proposed here provides a
new practical route to topological quantum computing,
mostly because we are not aware of potential realizations with
a built-in chiral symmetry, and because the set of available
quantum gates is very limited and hence not universal. 
Nevertheless, our proposal does have  two very appealing features. 
First, its single-particle nature is a major
conceptual simplification with respect to other systems 
showing topologically protected quantum dynamics, such 
as the toric code, Majorana qubits, or
fractional quantum Hall systems.
Second, this simplicity makes our model 
particularly feasible for experimental realization:
single-particle tight-binding models can be realized 
in various established platforms, such as
optical waveguide 
arrays\cite{SzameitReview,Martin,menssen2019photonic} and
cold atomic systems\cite{MeierSSH,MeierAnderson},
promising an alternative shortcut towards the experimental demonstration
of topologically protected quantum gates.

In what follows, we will assume that the reader
is familiar with the SSH model, serving as an elementary 
example of 1D chiral symmetric
topological insulators, and the related concepts
such as the fully dimerized limit of the model, 
chiral symmetry, time-reversal symmetry,
localized states at edges and domain walls, 
the trivial and topological phases of the SSH model, and its
topological invariant. 
This background is covered in Chapters 1 and 8.1 
of Ref.~\onlinecite{Asboth}.

%=========
\section{Moving a zero-energy mode localized at a domain 
wall in a fully dimerized SSH chain}
\label{sec:domainwall}
%=========

\begin{figure}
%\centering
%\hspace{-0.06\columnwidth}
\includegraphics[width=1.0\columnwidth]{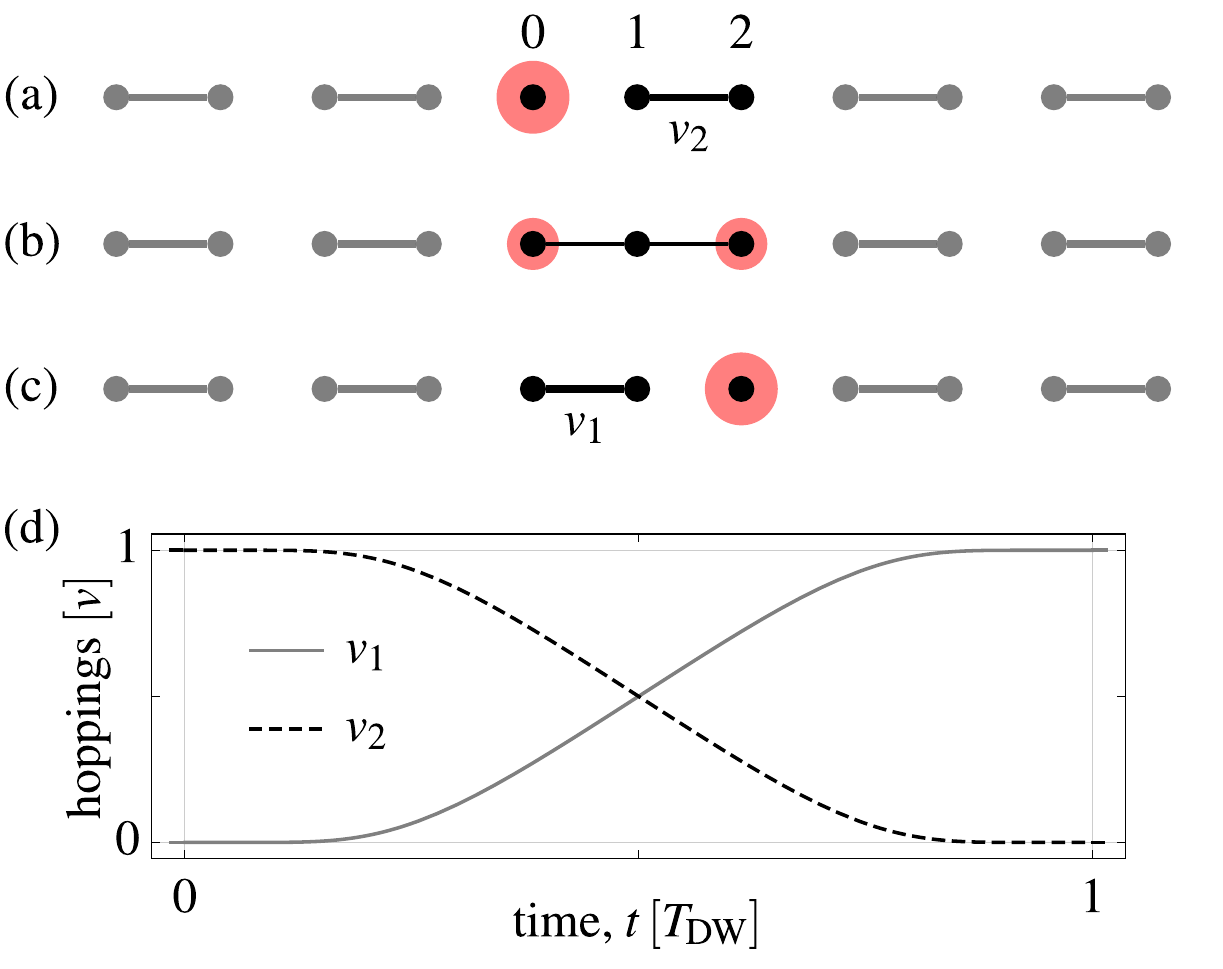}
\caption{Adiabatically moving a domain wall and 
the localized zero-energy state it supports, in 
a fully dimerized SSH chain.
Red circles depict the particle density of the localized state.
The sequence (a)-(b)-(c) shows how to move the
domain wall and the localized state by two sites.
The domain-wall movement time $T_\text{DW}$ is the
time window between (a) and (c).
(d) Time dependence of the hopping amplitudes,
see Eq.~\eqref{eq:timedependence}.
}
\label{fig:DomainWallStep}
\end{figure}

A well-known braiding protocol in a Majorana Y-junction\cite{Alicea}
is based on the adiabatic motion of defects, which 
are domain walls in that case. 
The braiding protocol in the SSH Y-junction, which we propose below
in section \ref{sec:dimerized},
is also based on moving domain walls.
In this section, we introduce a simple scheme to move a domain wall in
an SSH chain. A similar method is presented for topologically protected quantum state transfer in superconducting qubit chains \cite{FengQST}.

The elementary step of this domain-wall motion is illustrated in 
Fig.~\ref{fig:DomainWallStep}.
Figure \ref{fig:DomainWallStep}a shows a fully dimerized 
SSH chain with a domain wall that is the isolated site.
We consider the three-site black subset of the chain, 
labelled with 0, 1, and 2, 
and disregard the remaining gray part for our discussion. 
The Hamiltonian describing this three-site block reads
\bean
H(t) = v_1(t) \ket{1}\bra{0} + v_2(t) \ket{2} \bra{1} + h.c.,
\eean
with $v_1(t=0) = 0$ and $v_2(t=0) = v >0$ as shown 
in Fig.~\ref{fig:DomainWallStep}a.
Here, each ket denotes the state localized at the corresponding site.
We consider real-valued hopping amplitudes
in most of this work, unless noted otherwise.
It is a simple fact that the domain wall, 
i.e., site 0 in Fig.~\ref{fig:DomainWallStep}a, 
which is disconnected from the rest of the chain, supports
a perfectly localized zero-energy mode $\ket{0}$.
This state is depicted as the red circle. 

The domain wall together with this localized state can be moved 
adiabatically by, e.g., increasing the hopping $v_1$ to the value $v$, 
and decreasing the hopping $v_2$ to zero simultaneously.
Figures \ref{fig:DomainWallStep}a,b,c show the initial configuration, 
an intermediate configuration, and the final configuration, 
respectively. 
As a result of these changes, the domain wall has moved from site
0 (Fig.~\ref{fig:DomainWallStep}a) to site 2 (Fig.~\ref{fig:DomainWallStep}c).
Furthermore, if these hopping-amplitude changes 
are done adiabatically, then the
state $\ket{0}$ evolves to the state $- \ket{2}$, that is, 
the zero mode moves two sites to the right and acquires a minus sign. 
This is a direct consequence of the simple fact that
the Hamiltonian $H$ has a zero-energy eigenstate 
\bean
\label{eq:movingzeromode}
\ket{\psi} = 
\frac{v_2 \ket{0} - v_1 \ket{2}}{\sqrt{v_1^2+v_2^2}}.
\eean
The prefactor $-1$ in front of $\ket{2}$ above is not
supplemented by a 
dynamical phase factor: the dynamical phase vanishes 
because the state $\ket{\psi}$ is a zero-energy eigenstate
for all intermediate times.
Note that this instantaneous zero-energy state generally has
some weight on sites 0 and 2, but has no weight on 
site 1. 
For example, in the intermediate time step when 
$v_1 = v_2$, this zero-energy state is evenly distributed 
on the sites 0 and 2, as shown in Fig.~\ref{fig:DomainWallStep}b.

The actual time dependences of the hopping amplitudes $v_1$ 
and $v_2$ can have various forms, 
the resulting dynamics is independent of the details in 
the adiabatic limit. 
The simplest option could be to use  linear ramps for each
hopping amplitude.
In what follows, we will use a smooth, exponential 
time dependence instead, which is expected to suppress
leakage from the zero-energy subspace that arises due to 
the finite (non-infinite) time duration $T_\text{DW}$ 
of the domain-wall movement.\cite{Knapp,Sekania} 
The hopping amplitudes
are changed in time as
\begin{subequations}
\label{eq:timedependence}
\bean
v_1(t) &=& 
v \, \chi(t/T_\text{DW})
, \\
v_2(t) &=& v\left(1-\chi(t/T_\text{DW})\right),
\eean
\end{subequations}
where the pulse shape function $\chi$ is defined as
\bean
\chi(x) = \frac{e^{-1/x}}{e^{-1/(1-x)}+e^{-1/x}}.
\eean
In Eq.~\eqref{eq:timedependence}, 
we introduced the domain-wall movement time $T_\text{DW}$, 
which is the time used to move the domain wall by two sites
(i.e., from (a) to (c) in Fig.~\ref{fig:DomainWallStep}).
The time-dependent hopping amplitudes in Eq.~\eqref{eq:timedependence}
are shown in Fig.~\ref{fig:DomainWallStep}d.

Importantly, the result of this adiabatic deformation of the Hamiltonian,
i.e., that the state $\ket{0}$ develops to the state $-\ket{2}$, 
does not depend on the actual protocol used to tune 
the hopping amplitudes in time.
As seen directly from Eq.~\eqref{eq:movingzeromode},
this final state is guaranteed 
provided that at least one of $v_1$ 
and $v_2$ is nonzero for all times, and that the final value
of $v_1$ is positive.
%As long as chiral symmetry is maintained, the existence of the
%instantaneous zero-energy eigenstate is guaranteed, and
%the final state is always $-\ket{2}$. 
%In this case, the chiral symmetry operator is
%$\mathcal{C} = \ket{0}\bra{0} + \ket{2} \bra{2} - \ket{1} \bra{1}$, 
%and chiral symmetry is maintained as long as
%the on-site energies and 
%the next-nearest-neighbor hopping are zero. 

%=========
\section{Exchange of zero-energy modes in a Y-junction
provides a $Y$ gate}
%=========
\label{sec:dimerized}

In this section, we propose the SSH Y-junction setup and 
a braiding protocol, and numerically solve the time-dependent
Schr\"odinger equation to show that adiabatic braiding 
leads to a perfect $Y$ gate if
the SSH chains of the junction are fully dimerized. 
Topological protection is not addressed here, that
we will do in the subsequent sections.

\begin{figure}
%\centering
%\hspace{-0.06\columnwidth}
\includegraphics[width=1.0\columnwidth]{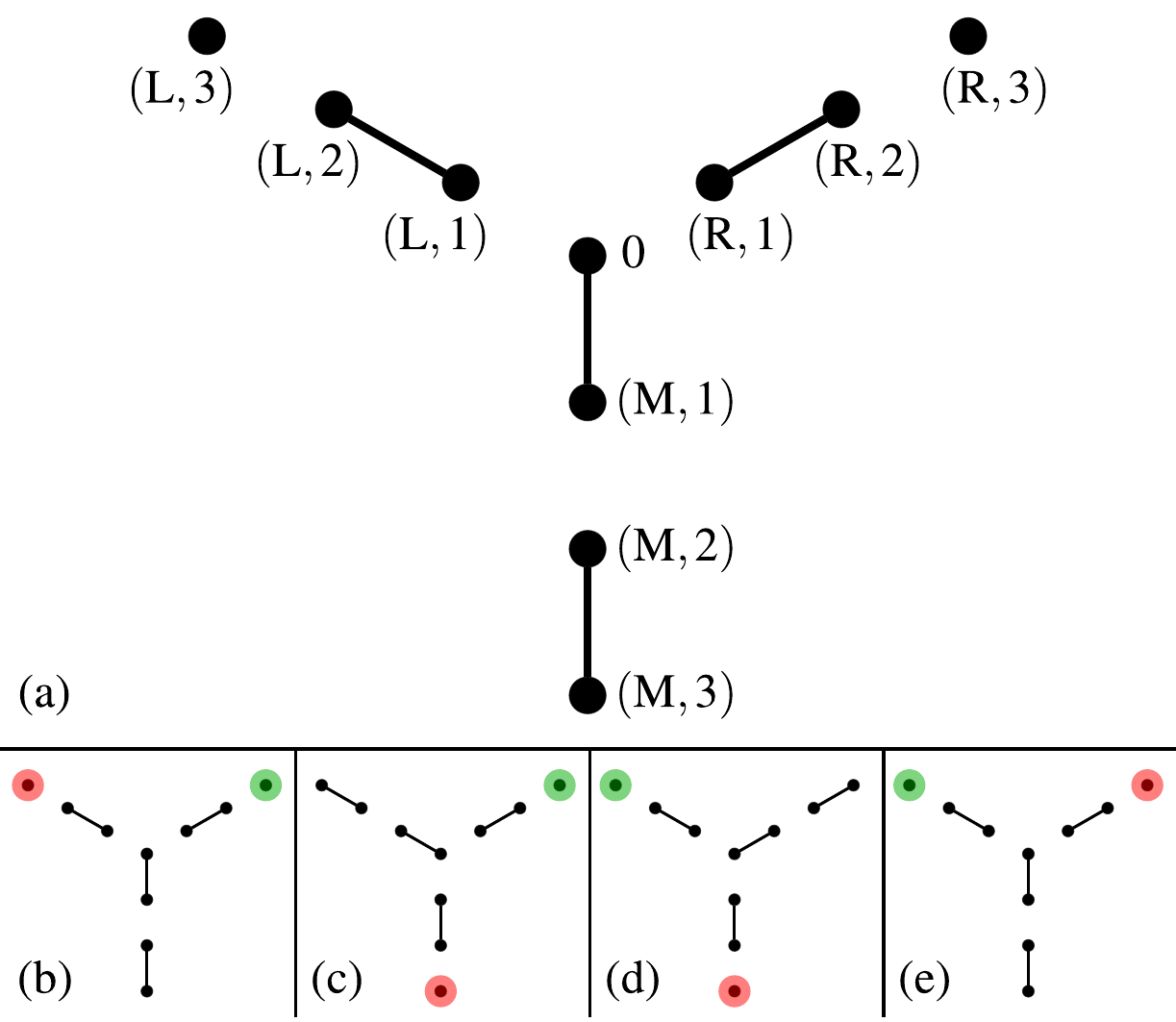}
\caption{Braiding in a Y-junction constructed from three fully dimerized SSH 
chains. 
(a) shows the initial configuration and the labels associated
to the sites. 
In (b)-(e), the colored sites (red, green) depict the zero-energy edge modes.
These edge modes are exchanged by adiabatically moving the domain
walls supporting them, following the scheme in 
Fig.~\ref{fig:DomainWallStep}.
(b) depicts the initial configuration, $t=0$.
(c) and (d) depicts the two intermediate configurations
when the zero-energy modes are localized at the outer ends of the chains, 
$t=3\, T_\text{DW}$ and $t=6\, T_\text{DW}$, respectively. 
(e) denotes the final configuration at
$t = 9\, T_\text{DW}$: the Hamiltonian is the same as the
initial one (b), but the edge modes have been exchanged.
}
\label{fig:DimerizedYJunction}
\end{figure}

The SSH Y-junction and the braiding protocol are 
illustrated in Fig.~\ref{fig:DimerizedYJunction}.
The single-particle tight-binding Hamiltonian
at the initial stage of the braiding protocol
is shown in Fig.~\ref{fig:DimerizedYJunction}a.
This junction is formed by three SSH chains, 
connected via a central site. 
We denote the three
chains by $\text{L}$, $\text{R}$ and $\text{M}$,
according to their location in the figure. 
For simplicity, we consider cases where the three chains
have the same length $N_c$; 
the figure corresponds to chain length $N_c=3$. 
The tight-binding Hamiltonian of this Y-junction reads
\bean
\nonumber
H &=& \sum_{c \in C}
\sum_{m = 1}^{N_c-1} v_{c,m} \left(\ket{c,m}\bra{c,m+1} + h.c. \right)\\
&+& 
\sum_{c \in C} v_{c,0} \left(\ket{0} \bra{c,1} + h.c. \right)
\label{eq:hamiltonian}
\eean
where $C = \{\text{L}, \text{R}, \text{M}\}$ is the set of 
chain indices, $m$ is the site index 
within a given chain. 
According to Fig.~\ref{fig:DimerizedYJunction}a, 
the state localized on the central site is denoted by $\ket{0}$, 
and the state localized on the $m$th site on chain $c$
is denoted by $\ket{c,m}$.

Figure \ref{fig:DimerizedYJunction}a depicts a certain 
configuration of hopping amplitudes: 
the hopping amplitudes shown as black lines are set to 
$v$ (e.g., $v_{\text{M,0}} = v$), 
all other hopping amplitudes are set to zero
(e.g., $v_{\text{M,1}}=0$). 
The Hamiltonian corresponding to 
Fig.~\ref{fig:DimerizedYJunction}a
has a two-dimensional 
zero-energy subspace, spanned by the states
$\ket{\text{L},3}$ (depicted as the red circle in 
Fig.~\ref{fig:DimerizedYJunction}b), 
and
$\ket{\text{R},3}$ (depicted as the green circle in
Fig.~\ref{fig:DimerizedYJunction}b).
These two edge sites can be considered as defects giving
rise to a two-dimensional zero-energy subspace.
 
We propose to perform the $Y$ gate in
this two-dimensional subspace by 
exchanging 
the two defects adiabatically. 
The first stage of the 
exchange consists of  moving the red defect by
repeating the elementary step 
introduced in Fig.~\ref{fig:DomainWallStep}
(i.e., moving the domain wall by two sites)
three times, to achieve the configuration in 
Fig.~\ref{fig:DimerizedYJunction}c;
the second and third stages are analogous, 
yielding the configurations in Fig.~\ref{fig:DimerizedYJunction}d
and e, respectively. 
After the third stage, the Hamiltonian returns to its initial
form, but the two defects and the two localized zero-energy
modes have been exchanged. 
Actually, since it took an odd number of steps (three) to move
the green defect, the state $\ket{\text{R},3}$ evolves
into the state $-\ket{\text{L},3}$, picking up a minus sign. 
In contrast, it took an even number of steps (six) 
to move the red defect, hence the state $\ket{\text{L},3}$ evolves
into the state $\ket{\text{R},3}$, with no sign change. 
As a result, the braiding induces a gate 
\bean
\label{eq:ygate}
Y  = \ttmatrix{0}{-1}{1}{0} = - i\sigma_y 
\eean
in the basis 
$(\ket{\phi_1},\ket{\phi_2}) \equiv (\ket{\text{L},3},\ket{\text{R},3})$. 
It is clear that
a similar braiding operation in a Y-junction with
an arbitrary odd chain length $N_c=1,3,5,\dots$ 
yields the same gate. 
Note that the elementary version of this scheme, corresponding to 
$N_c = 1$, was studied in Refs.~\onlinecite{Unanyan,Droth}.

The braiding should provide an exact $Y$ gate
in the adiabatic limit, i.e., for infinitely long braiding time $T$.
In the rest of this section, we study if that is true, and 
how the finite braiding time
affects the accuracy of the gate.
We characterize the accuracy
using the concept of average fidelity. 
The time evolution during the 
full braiding sequence is described by the
propagator of the time-dependent Schr\"odinger equation,
that is expressed by the well-known time-ordered ($\mathcal{T}$)
exponential
\bean
U(T) = \mathcal{T} \exp\left(
- \frac{i}{\hbar} \int_0^T dt H(t)
\right).
\eean 
The effect of the braiding on the two-dimensional zero-energy
subspace is described by the propagator projected 
onto that subspace, which we will refer to as the 
\emph{overlap matrix}:
\bean
\label{eq:overlapmatrix}
\mathcal{O}(T) = \ttmatrix{\braket{\phi_1 | U(T) | \phi_1}}{\braket{\phi_1 | U(T) | \phi_2}}{\braket{\phi_2 | U(T) | \phi_1}}{\braket{\phi_2 | U(T) | \phi_2}}.
\eean
As stated above, we expect that braiding is perfect in the
adiabatic limit, that is, $\lim_{T \to \infty} \mathcal{O}(T) = Y$.

For a given initial state $\ket{\phi}$ in the two-dimensional 
zero-energy subspace, the fidelity of the braiding operation 
can be described by the probability of finding the
state after the braiding operation in the final state
of the idealized operation: 
$f(\phi,T) = \left| \braket{\phi | Y^\dag \mathcal{O}(T) | \phi}\right|^2$.
Therefore, the overall quality of the gate can be described by
the average of the above fidelity 
$f(\phi,T)$ 
for all initial states $\phi$ in the
two-dimensional zero-energy subspace, yielding\cite{Pedersen}
\bean
F = \frac 1 6 \left[
	\text{Tr} \left(\mathcal{O}^\dag \mathcal{O} \right)
	+
	\left|
		\text{Tr} \left(Y^\dag \mathcal{O}\right)
	\right|^2
\right].
\label{eq:avgfid}
\eean
where the argument $T$ of $F(T)$ and $\mathcal{O}(T)$ was omitted
for brevity. 
The \emph{average fidelity} is $F=1$
if the overlap matrix  $\mathcal{O}$ is 
equivalent to the ideal gate $Y$, and
$0 \leq F < 1 $
otherwise.
Hence, we characterize the error of the braiding protocol 
using the \emph{infidelity} 
\bean
\label{eq:infidelity}
\varepsilon(T) = 1- F(T).
\eean

\begin{figure}
%\centering
%\hspace{-0.06\columnwidth}
\includegraphics[width=1.0\columnwidth]{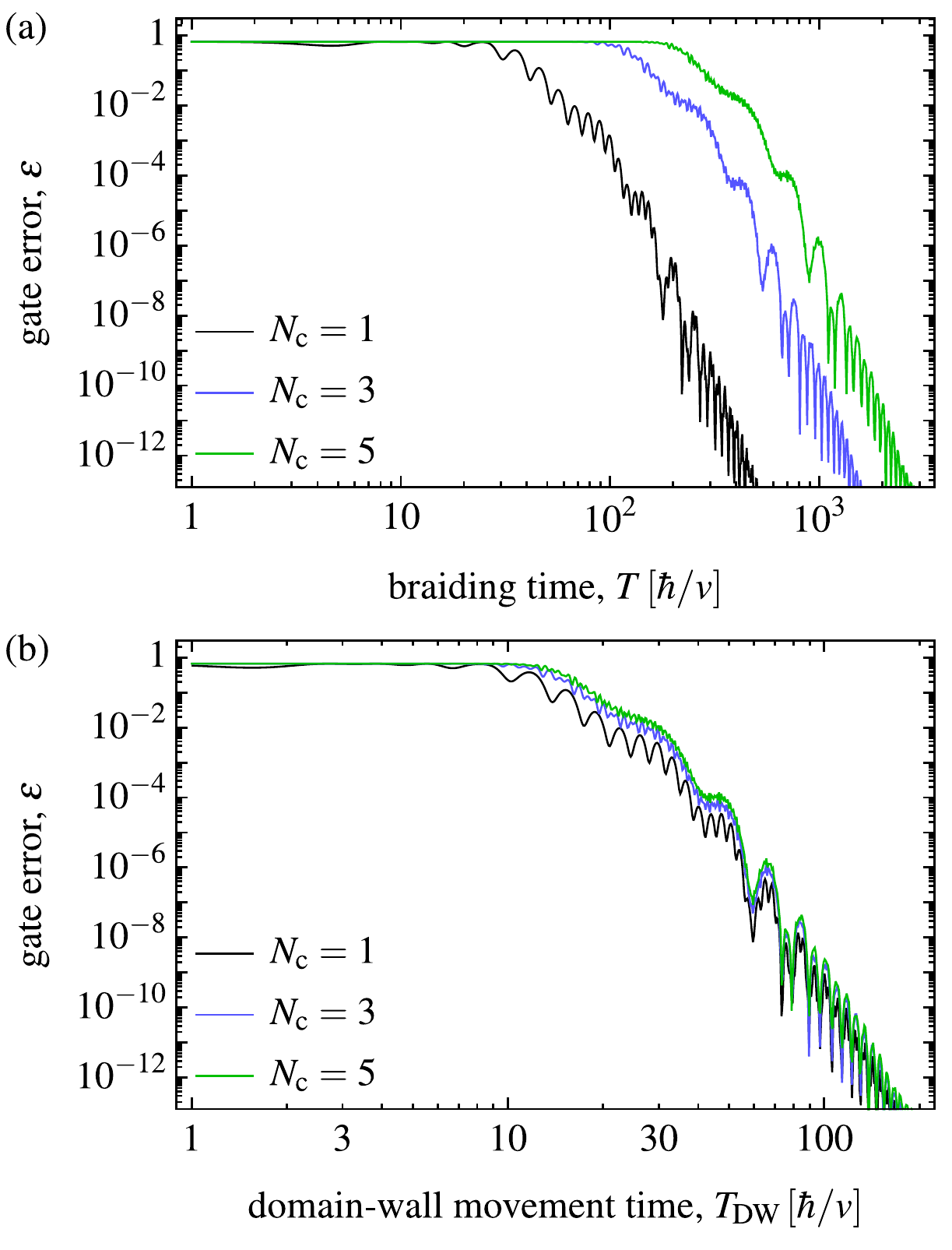}
\caption{Error of the $Y$ gate in the fully
dimerized SSH Y-junction due to the finite
braiding time.
(a) Gate
error (infidelity, see Eq.~\eqref{eq:infidelity}) 
is shown as the function of braiding time.
(b) Gate error of panel (a) is shown, rescaled, 
as the function of the time $T_\text{DW}$
required for a single step
of domain-wall motion. 
Errors are induced due to the non-adiabatic character of the braiding, 
hence they get suppressed as the braiding time is increased.  
}
\label{fig:DimerizedResult}
\end{figure}

The error $\varepsilon$, obtained from a numerical solution of the 
time-dependent Schr\"odinger equation, 
is shown in Fig.~\ref{fig:DimerizedResult}a, as a function 
of braiding time $T$, for three different lengths $N_c = 1,3,5$ of the chains
forming the Y-junction.
For short braiding times, errors are caused by the non-adiabatic
character of the driving. 
Even though there are oscillations on each curves, 
the error seems to converge to zero as the braiding time 
is increased, confirming expectations.
Furthermore, the 
results suggest that it is possible to reach any targeted error level,
but the longer the chain, the longer braiding time is required for that.
Finally, an interesting scaling property is demonstrated in  
Fig.~\ref{fig:DimerizedResult}b, where the data set is the same as in 
the Fig.~\ref{fig:DimerizedResult}a, 
but the horizontal axis is rescaled to show
the domain-wall movement time $T_\text{DW}$
instead of the braiding time. 
With this scaling, the three data sets show a very similar
behavior, indicating that the velocity of the domain-wall motion 
is the factor determining the gate error.

Up to now, we have studied a simple protocol that performs
a quantum gate in a degenerate subspace of a Hamiltonian. 
Is this gate robust in any sense?
Is it robust if we relax the fully dimerized character of the Y-junction
that was assumed in this section?
Is it robust if we introduce disorder? 
We address these questions in what follows.

%=========
\section{Numerical demonstration of topological protection}
\label{sec:nondimerized}
%=========

In the previous section, we assumed that the SSH Y-junction
is in a fully dimerized configuration, apart from the region of the
domain wall that is being moved. 
Here, we consider a case when the system is not 
in the fully dimerized configuration, and in addition, 
random hopping disorder
with real-valued hopping amplitudes  
is also introduced. 
In this case, chiral symmetry 
and the time-reversal symmetry of the setup are still preserved. 
We show that the $Y$ gate is topologically protected, i.e., 
becomes perfect, insensitively to the disorder, in the adiabatic limit
and 
in the limit of large system size (Fig.~\ref{fig:HoppingDisorder}). 
In contrast, as we also show below,
there is no such protection, when disorder on the on-site energies
of the lattice 
is introduced
 (Fig.~\ref{fig:OnsiteDisorder}), 
 or when the hopping disorder consists of complex-valued
 hopping amplitudes (Fig.~\ref{fig:ComplexDisorder}).

\subsection{The $Y$ gate is robust against real-valued hopping disorder}

\begin{figure}
%\centering
%\hspace{-0.06\columnwidth}
\includegraphics[width=1.0\columnwidth]{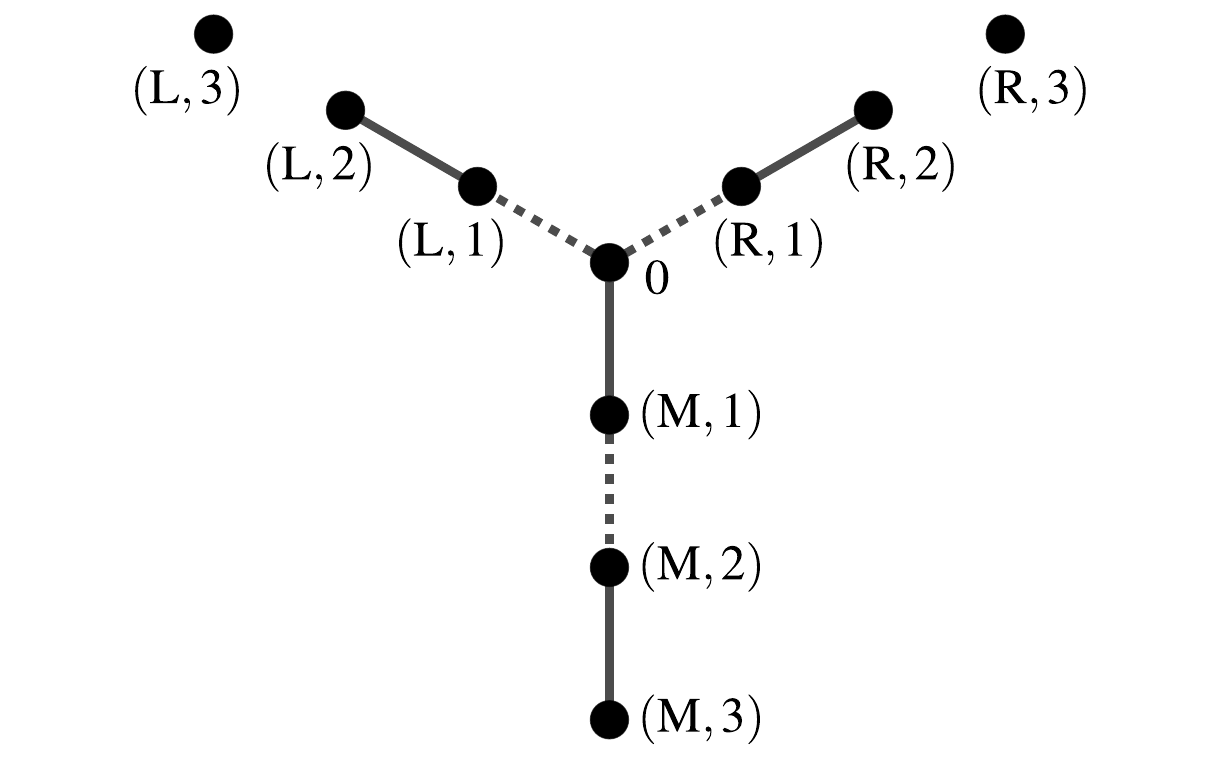}
\caption{SSH Y-junction away from the fully dimerized limit with
hopping disorder.
Structure of the initial and final Hamiltonian
of the braiding protocol is depicted. 
Solid (dashed) lines are strong (weak) hopping amplitudes, all
having random components. 
In the initial and final configuration shown here, 
the defects residing on sites $(\text{L},3)$ and $(\text{R},3)$ are isolated from the rest of the
system.}
\label{fig:NonDimerizedYJunction}
\end{figure}

We consider a setup similar to Fig.~\ref{fig:DimerizedYJunction}a,
see Fig.~\ref{fig:NonDimerizedYJunction}.
The setup in Fig.~\ref{fig:NonDimerizedYJunction} is also 
described by the Hamiltonian \eqref{eq:hamiltonian}.
Similarly to the case of Fig.~\ref{fig:DimerizedYJunction}a, 
we keep all on-site energies zero, and use real-valued time-dependent
hopping amplitudes. 
In this case, however, the hopping amplitudes are not tuned
between 0 and $v$, but between a minimal value 
$v^\text{min}_{c,m}$ and a maximal value
$v^\text{max}_{c,m}$ instead.
As indicated in Fig.~\ref{fig:NonDimerizedYJunction},
the defect sites are isolated,
i.e., we choose 
$v^\text{min}_{\text{L},2} = v^\text{min}_{\text{R},2} = 0$
and $v^\text{max}_{\text{L,2}} = v^\text{max}_{\text{R},2} = v$. 
However, we introduce  disorder 
in all further hopping amplitudes depicted in 
Fig.~\ref{fig:NonDimerizedYJunction}:
\begin{subequations}
\label{eq:randomhopping}
\bean
v_{c,m}^\text{min} &=& w + \delta v_{c,m}, \\
v_{c,m}^\text{max} &=& v + \delta v_{c,m}.
\eean
\end{subequations}
Here, $w < v$ is chosen, and $\delta v_{c,m}$ is
a Gaussian random variable with zero mean 
and standard deviation $s_v < v$. A finite (non-zero) value of $w$ corresponds to the non-fully dimerized case when the defect-bound states spread out and be more delocalized compared to fully dimerized limit. Same behavior is caused by the hopping disorder in a random fashion.

Apart from this difference in the minimal and maximal
values of the hopping amplitudes, the braiding
scheme is defined in complete analogy with
that in section \ref{sec:dimerized}.
For example, the first elementary step 
in moving the red zero mode and the 
associated defect is induced by
\begin{subequations}
\label{eq:protocol}
\bean
v_{\text{L},2}(t) &=& v \, \chi(t/T_\text{DW}) ,\\
v_{\text{L},1}(t) &=& v+ \delta v_{\text{L},1}-(v-w)
\chi(t/T_\text{DW}). 
\eean
\end{subequations}

\begin{figure}
%\centering
%\hspace{-0.06\columnwidth}
\includegraphics[width=1.0\columnwidth]{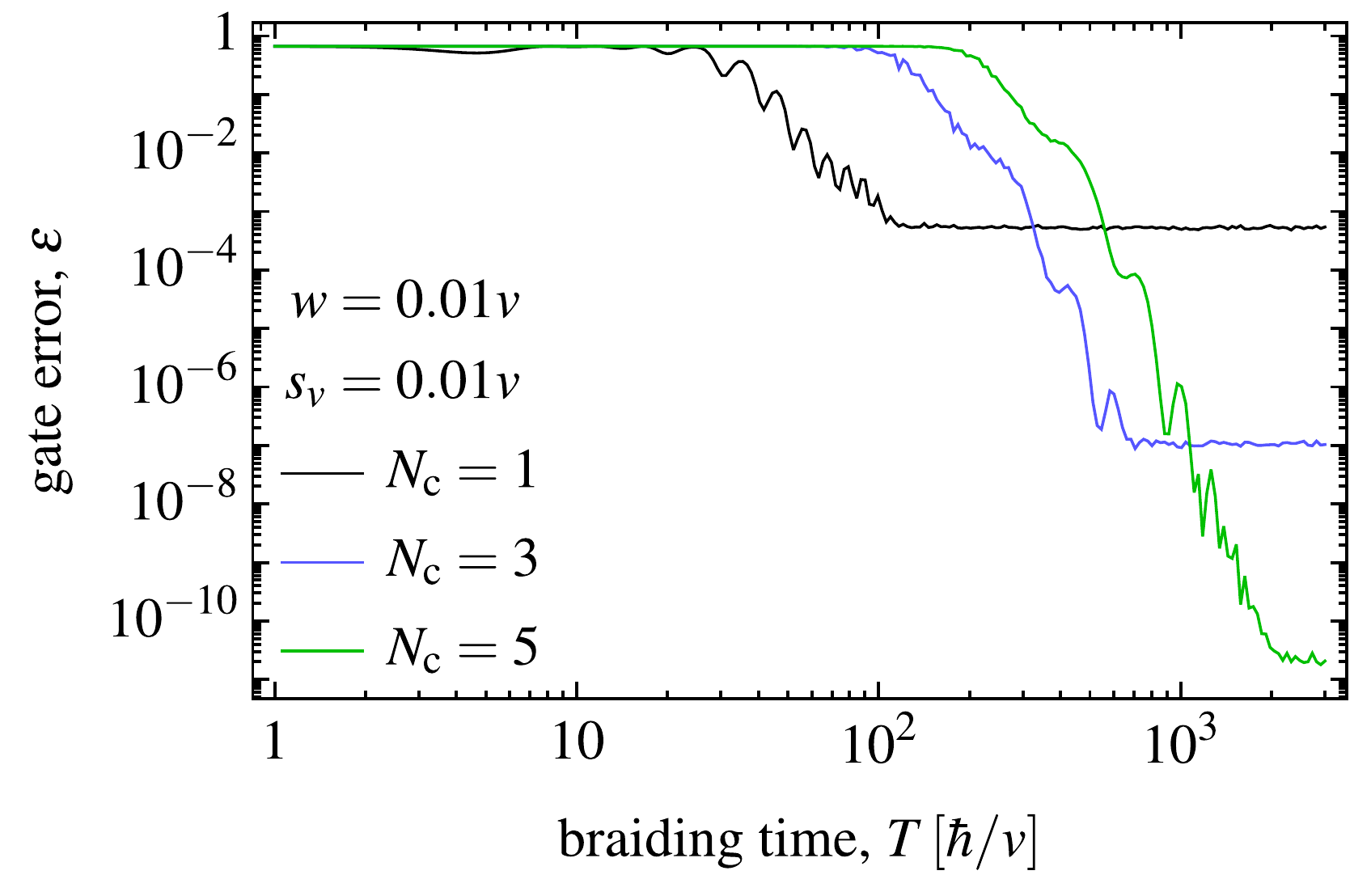}
\caption{Error of the $Y$ gate in a non-dimerized
SSH Y-junction with real-valued hopping disorder.
Disorder-averaged 
gate error (infidelity, see Eq.~\eqref{eq:infidelity}) is shown 
as a function of braiding time $T$ for three different chain lengths
$N_c$.
The error saturates for long braiding times in each case, 
and the error minimum decreases exponentially 
as the chain length is increased, despite the presence of disorder;
this indicates topological protection. 
Each curve is an average for 1000 random realizations.
}
\label{fig:HoppingDisorder}
\end{figure}

The robustness of the $Y$ gate, i.e., its resilience to 
the real-valued 
hopping disorder in the adiabatic and long-chain limit is
shown in Fig.~\ref{fig:HoppingDisorder}.
The zero-energy subspace of the initial Hamiltonian
(which equals the final Hamiltonian) is spanned by the 
two completely localized states $\ket{\text{L},3}$ and
$\ket{\text{R},3}$, and therefore we can characterize
the gate fidelity using the  infidelity $\varepsilon = 1-F$ introduced earlier 
in Eq.~\eqref{eq:infidelity}.
In Fig.~\ref{fig:HoppingDisorder}, we show the error $\varepsilon$
of the braiding-based $Y$ gate as the function of braiding
time $T$, for three different chain lengths $N_c$.
Furthermore, we set $w=0.01v$ and the disorder strength
to $s_v = 0.01v$, 
and each curve is an average for 1000 different disorder realizations.
The first key feature of these results is that 
the gate error approaches its minimum value for long 
braiding times, when braiding is adiabatic, as expected. 
In fact, the error saturates and forms a plateau in all three cases. 
The second key feature is that the minimum value of the error, i.e., 
the height of the plateau, decreases exponentially
as the chain length is increased.
Here we do not show, but a finite $w$ or a hopping disorder with a finite $s_v$ leads separately to the saturation of the error.
These results clearly demonstrate the robustness
of the $Y$ gate against hopping disorder even away from the fully dimerized limit.
As we show in section \ref{sec:topologicalprotection}, this resilience can be 
understood as topological protection, 
arising as a joint consequence  of chiral symmetry, 
time-reversal symmetry
(both of which are preserved in the presence of 
real-valued hopping disorder), 
and the spatial separation of the zero modes during their
exchange. 

\subsection{The $Y$ gate is not robust against 
on-site disorder}

\begin{figure}
%\centering
%\hspace{-0.06\columnwidth}
\includegraphics[width=1.0\columnwidth]{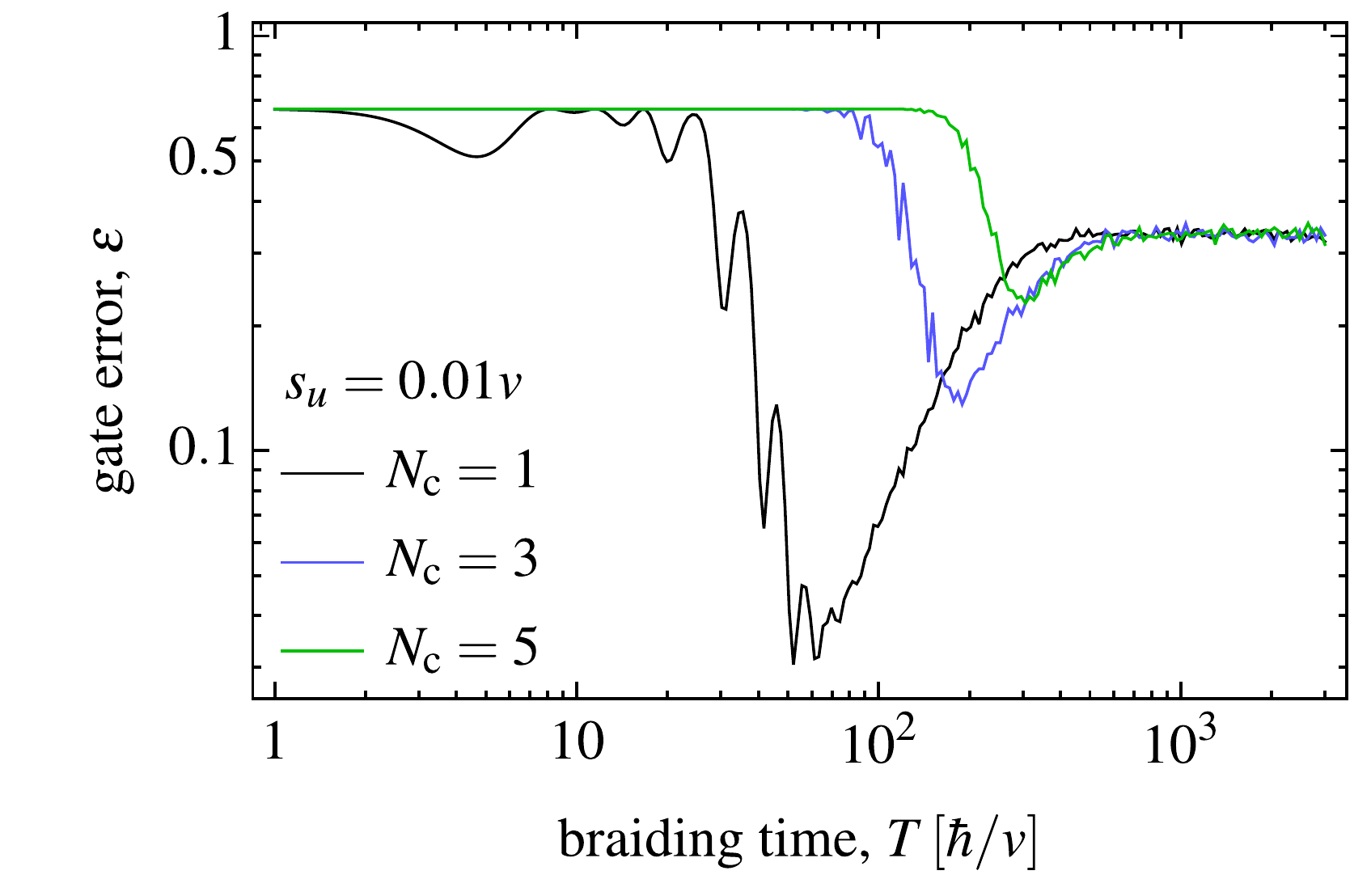}
\caption{Error of the $Y$ gate in the presence
of on-site disorder.
Disorder-averaged gate error (infidelity, see Eq.~\eqref{eq:infidelity})
is shown as a function of braiding time $T$ for three different
chain lengths $N_c$.
The $Y$ gate is not robust against on-site disorder:
the gate error saturates at a high value for long braiding times, 
at a value independent of the chain length.
Each curve is an average for 1000 random realizations.}
\label{fig:OnsiteDisorder}
\end{figure}

The braiding-based $Y$ gate is not resilient to 
on-site energy disorder.
This is illustrated by Fig.~\ref{fig:OnsiteDisorder}.
For simplicity, in this subsection 
we switch off hopping disorder ($s_v = 0$)
and take $w=0$. 
To study the effect of on-site disorder, we supplement the Y-junction 
Hamiltonian \eqref{eq:hamiltonian} with the on-site term.
That is, we consider dynamics governed by the 
Hamiltonian $H' = H + H_\text{onsite}$, 
where the second term is defined as 
\bean
\label{eq:onsite}
H_\text{onsite} = \sum_{j} u_j \ket{j} \bra{j}.
\eean
Here, the  parameters $u_j$ are random on-site energies 
with zero mean, and, in what follows, with a 
standard deviation $s_u = 0.01 v$.
The sum in Eq.~\eqref{eq:onsite}
goes for all sites except the two sites supporting the two defects
at $t=0$, that is, except $(\text{L},3)$ and $(\text{R},3)$.

The numerically obtained gate error as a function of braiding time
is shown in 
Fig.~\ref{fig:OnsiteDisorder} for three different chain lengths.
To obtain this data, we have numerically solved the time-dependent
Schr\"odinger equation of the Hamiltonian $H'$ defined above.
Each curve in 
Fig.~\ref{fig:OnsiteDisorder}
is an average for
1000 random disorder realizations.
The key differences in comparison to the 
case with real-valued hopping disorder in Fig.~\ref{fig:HoppingDisorder} 
are as follows.
(i) The gate error is always much larger
in the presence of on-site disorder than in the presence
of real-valued hopping disorder.
(ii) For real-valued hopping disorder, 
the minimal error gets smaller for longer 
chain length; for on-site disorder the trend is the opposite. 
(iii) For on-site disorder, the error does not have a minimum in the
adiabatic, long-braiding-time limit, but 
at an intermediate braiding time. 
In short, a key reason behind these differences is that
the on-site disorder opens a minigap 
in the spectrum between the two zero-energy eigenstates, 
the corresponding dynamical phase picked up by the states
during the time evolution depends explicitly on the random disorder
configuration and the braiding time, and hence it leads to 
dephasing in the adiabatic limit.

\subsection{The $Y$ gate is not robust against 
complex-valued hopping disorder}

\begin{figure}
%\centering
%\hspace{-0.06\columnwidth}
\includegraphics[width=1.0\columnwidth]{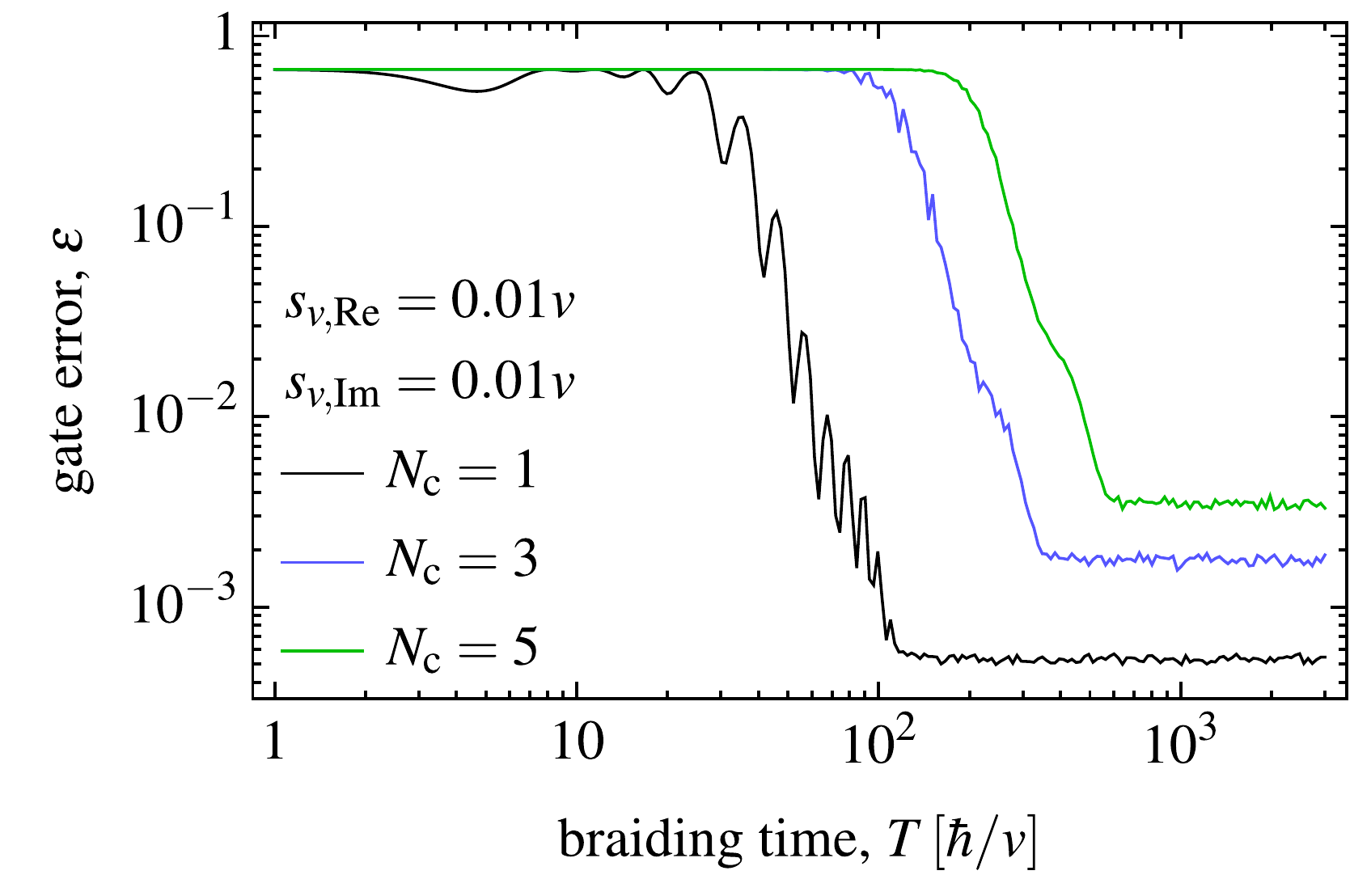}
\caption{Error of the $Y$ gate in the presence
of complex-valued hopping disorder. 
Disorder-averaged gate error (infidelity, see Eq.~\eqref{eq:infidelity})
is shown as a function of braiding time $T$ for three different
chain lengths $N_c$.
The error saturates at a value for long braiding times, 
but that value increases as the chain length is increased.
This indicates that the $Y$ gate is not robust
against complex-valued hopping disorder.
Each curve is an average of 1000 random realizations.}
\label{fig:ComplexDisorder}
\end{figure}  

The braiding-based $Y$ gate is not resilient to 
hopping disorder, if the hopping amplitudes can take complex values
(Fig.~\ref{fig:ComplexDisorder}).
To show this, 
we use the same parametrization of the hopping amplitudes
as introduced in Eqs.~\eqref{eq:randomhopping} and \eqref{eq:protocol}, 
with the only difference that the random 
components $\delta v_{c,m}$ 
of the hopping amplitudes are complex.
In particular, we consider the case
when the real and imaginary parts of 
$\delta v_{c,m}$ are identically distributed normal 
random variables, with standard deviations
$s_{v,\text{Re}} = s_{v,\text{Im}} = 0.01 v$. 
Furthermore, we set $w=0$ for simplicity. 

The numerically obtained gate error as a function of braiding time
is shown in Fig.~\ref{fig:ComplexDisorder}
for three different chain lengths. 
The key feature of the results is that the
error saturates at a plateau for each chain length,
similarly to the case of real-valued hopping disorder,  
but the height of the plateau actually grows as the chain length is
increased.
This is in contrast to the case of real-valued hopping
disorder, where we have seen that the error plateau height
decreases exponentially as the chain length is increased. 
Hence, the results in Fig.~\ref{fig:ComplexDisorder} indicate
that the $Y$ gate is not protected against
complex-valued disorder.

%=======
\section{Topological protection is ensured by chiral 
symmetry, time-reversal symmetry, 
and spatial separation of the zero modes}
\label{sec:topologicalprotection}
%=======

In the previous section, we have demonstrated
that the braiding-based $Y$ gate is robust against
real-valued hopping disorder, but not robust against 
on-site disorder, neither against complex-valued hopping disorder. 
Here, we prove that this robustness is due to topological protection, 
and arises as a joint consequence of
chiral symmetry, time-reversal symmetry, 
and the spatial separation of the
defects supporting the zero-energy modes.

As the first step of the proof, we  introduce two sublattices: 
sublattice $A$ contains the sites with an odd ordinal number,
e.g., $(\text{L},3)$, $(\text{C},1)$, etc, 
whereas sublattice $B$ contains the rest of the sites. 
Then, the matrix 
\bean
\mathcal{C} = \sum_{j \in A} \ket{j} \bra{j}
-\sum_{j \in B} \ket{j} \bra{j}
\eean
is a chiral symmetry\cite{Asboth} 
of the Hamiltonian, that is,
$\mathcal{C} H \mathcal{C}^{-1} = - H$.
This is due to the fact that all matrix elements of the
Hamiltonian are connecting sites of different sublattices.
Note that there are more $A$ sites than $B$ sites:
we have  $N_A = N_B + 2$, where 
$N_A$ ($N_B$) is the number of $A$ ($B$) sites;
in the example of Fig.~\ref{fig:DimerizedYJunction}a, 
$N_A = 6$ and $N_B=4$. 
As a consequence of this mismatch and the chiral symmetry, 
the Hamiltonian has two degenerate zero-energy eigenstates, 
as we prove in Appendix \ref{app:darkstatetheorem}.
This zero-energy subspace is different from the 
one formed by the edge states of a single SSH chain 
in the topological phase\cite{Asboth}: in the latter case, 
the edge states can hybridize and a nonzero minigap can
open between the bonding and antibonding hybrid states, 
whereas the defect-bound states in our Y junction
are at exactly zero energy, as a consequence of their dark-state character (see Appendix \ref{app:darkstatetheorem}).
 
 Our second step is to show that the overlap matrix $\mathcal{O}$
 defined in Eq.~\eqref{eq:overlapmatrix} is real-valued if the Hamiltonian,
 during the whole duration of the braiding, maintains
 its chiral symmetry and real-valuedness.
 To see this, 
 we factorize the propagator $U$ in $n+1$
 discrete short time steps of duration
$\tau = T / (n+1)$, 
that is, $U = U_n U_{n-1} \dots U_1 U_0$, 
where $U_j = \exp\left(- i H(j \tau) \tau/\hbar\right)$.
With this discrete representation of the propagator, we find 
\begin{subequations}
\bean
	\label{eq:realnessproof1}
	\mathcal{O}_{ij} & = &
	\braket{\phi_i |U | \phi_j} \\
		\label{eq:realnessproof2}
	&=&
	\braket{\phi_i |
	\mathcal{C}
	\left(
		\mathcal{C}
		U_n
		\mathcal{C}
	\right)
	\dots
		\left(
		\mathcal{C}
		U_0
		\mathcal{C}
	\right)
	\mathcal{C}
	| \phi_j}
	\\
	&=&
	\braket{\phi_i |
	\mathcal{C}
		U^*_n
	\dots
		U^*_0
		\mathcal{C}
	|\phi_j} 
			\label{eq:realnessproof3}
	\\
	&=& 
	\braket{\phi_i^* |
		U^*_n
	\dots
		U^*_0
	|\phi_j^*}
	 = \mathcal{O}_{ij}^*.
		\label{eq:realnessproof4}
\eean
	\label{eq:realnessproof}
\end{subequations}
To obtain \eqref{eq:realnessproof2} from \eqref{eq:realnessproof1}, 
we inserted unity in the form of $\mathcal{C}^2 = 1$ before and
after each time step.
To obtain \eqref{eq:realnessproof3} from \eqref{eq:realnessproof2},
we used the chiral symmetry of the Hamiltonian as well as
its real-valued character,
which imply 
\begin{subequations}
\bean
\mathcal{C} U_j \mathcal{C} &=&
\mathcal{C} \exp\left(- i H(j \tau) \tau/\hbar\right) \mathcal{C} \\
&=&
\exp\left(i H(j \tau) \tau/\hbar\right) \\
&=&
\exp\left(i H^*(j \tau) \tau/\hbar\right) = U^*_j.
\eean
\end{subequations}
To obtain \eqref{eq:realnessproof4} from \eqref{eq:realnessproof3}, 
we used the fact that   
the basis vectors
$\ket{\phi_i}$ are their own chiral partners, 
$\mathcal{C}\ket{\phi_i} = \ket{\phi_i}$
and that they are real-valued.

As the third step, we argue that in the limit of adiabatic 
braiding, the overlap matrix $\mathcal{O}$ is not only
real-valued but also unitary. 
To see that, we have to assume 
that during the braiding procedure, the 
zero-energy subspace of the Hamiltonian $H(t)$ 
is exactly twofold degenerate for all times, that is, 
that no lower-lying or higher-lying energy level ever
collides with the two zero-energy levels we focus on. 
Then, the notion of adiabaticity is indeed well-defined, 
and there is no leakage from the zero-energy subspace
if the braiding is adiabatic, guaranteeing the unitary 
nature of the overlap matrix.
All real-valued $2\times 2$ unitary matrices (i.e., the $2\times 2$ orthogonal matrices) can be written in one
of these forms:
\begin{subequations}
\bean
\mathcal{O}_+(\theta) &=& \left(\begin{array}{cc}
\cos \theta &  \sin \theta \\
- \sin \theta &  \cos\theta
\end{array} \right), \\
\mathcal{O}_-(\theta) &=& \left(\begin{array}{cc}
\cos \theta &  \sin \theta \\
\sin \theta &  -\cos\theta
\end{array} \right),
\eean
\end{subequations}
where $\theta$ is a real angle. 
One difference between these two classes is 
their determinant being $+1$ and $-1$, respectively.

In the last, fourth step, we
utilize that the two localized zero-energy 
eigenstates are spatially exchanged during the braiding
protocol.
As long as their spatial separation is ensured by using long
enough chains, 
the braiding procedure will move the first zero mode to 
the position of the second one, and move the second one
to the position of the first one. 
Therefore, the overlap matrix is off-diagonal, allowing only
four different configurations:
\begin{subequations}
\label{eq:braidingmatrices}
\bean
\mathcal{O}_{+}(\pi/2) &=& \left(\begin{array}{cc}
0 &  1 \\
-1 &  0
\end{array} \right) = i\sigma_y, \\
\label{eq:ygate2}
\mathcal{O}_{+}(-\pi/2) &=& \left(\begin{array}{cc}
0 &  -1 \\
1 &  0
\end{array} \right) = -i\sigma_y, \\
\mathcal{O}_{-}(\pi/2) &=& \left(\begin{array}{cc}
0 &  1 \\
1 &  0
\end{array} \right) =\sigma_x, \\
\mathcal{O}_{-}(-\pi/2) &=& \left(\begin{array}{cc}
0 &  -1 \\
-1 &  0
\end{array} \right) = -\sigma_x.
\eean
\end{subequations}

In conclusion, the characteristics of our braiding procedure 
restrict the overlap matrix of the braiding operation 
to the four options shown in Eq.~\eqref{eq:braidingmatrices}.
We know that in the fully dimerized limit, we have a $Y$ gate,
see Eq.~\eqref{eq:ygate}, which corresponds to  
Eq.~\eqref{eq:ygate2}.
Therefore, in the case of hopping disorder, when the Hamiltonian
is adiabatically connected to the fully dimerized limit, 
the overlap matrix in the adiabatic limit and limit
of long chains must be $Y$ as well.
We emphasize that we have used the
chiral symmetry and the real-valued nature of the
Hamiltonian to arrive to this conclusion. 
In what follows, we will refer to the real-valued nature 
of the Hamiltonian by saying that $H$ has time-reversal symmetry. 
In this case, time reversal symmetry $T$ is represented by 
the complex conjugation, $T=K$, this time-reversal
symmetry squares to $T^2 = 1$, and 
the real-valued nature of $H$ can be written as $KHK^{-1} = H$.
Hence, our system falls into the Cartan symmetry class
BDI\cite{ChingKaiChiu}.

A follow-up question is if the other three overlap
matrices $i\sigma_y$, $\sigma_x$, $-\sigma_x$ can be realized
with braiding-like processes. 
The answer is yes for $i\sigma_y$: 
one can achieve this by reversing the steps shown in 
Fig.~\ref{fig:DimerizedYJunction}b-e, 
by moving the right zero mode first.

Consider now further adiabatic deformations of the Hamiltonian,
where each mode ends up at its original position. 
Arguments similar to those used above imply that 
in such cases, the overlap matrix is one
of these four diagonal, real unitary $2\times 2$ matrices:
$1$, $\sigma_z$, $-1$, $-\sigma_z$.
It is clear that in our setup, $1$ can be realized by any 
adiabatic deformation of the Hamiltonian that does not braid
the zero-mode positions, 
and $-1 = (-i\sigma_y)(-i\sigma_y) = (i\sigma_y)(i\sigma_y)$ 
can be realized, e.g., by repeating the 
same braiding twice.
Below, we show that the remaining gates $\sigma_x$, $-\sigma_x$,
$\sigma_z$ and $-\sigma_z$ can not be realized.

All the quantum gates discussed above are topologically 
protected in the sense that their form depends
only on the topology of the world lines of the zero modes.
To formalize that a bit more, 
let us discuss arbitrary cyclic adiabatic deformations
that start and end with two isolated zero modes
(as done in the previous section), and where the spatial 
separation of the zero modes is guaranteed throughout
the deformation. 
Then, the world lines of the two zero modes for a given cyclic
deformation are characterized by a certain element
of the braid group $B_2$, which is equivalent to
the group of integer numbers $\mathbb{Z}$ with addition
as the group operation: a clockwise exchange contributes $+1$, 
an anticlockwise exchange contributes $-1$.
The set of the four quantum gates $1$, $Y$, $Y^2=-1$, $Y^3=-Y$, 
equipped with 
matrix multiplication as the group operation, form
a group $G_Y$ 
equivalent to the cyclic group of order 4, that is, $\mathbb{Z}_4$.
Our model therefore provides 
a two-dimensional representation $\rho$ of the braid group $B_2$ 
in terms of the quantum gates, 
namely $\rho:\,  B_2 \to G_Y, \, n \mapsto Y^n$.

\begin{figure}
%\centering
%\hspace{-0.06\columnwidth}
\includegraphics[width=1.0\columnwidth]{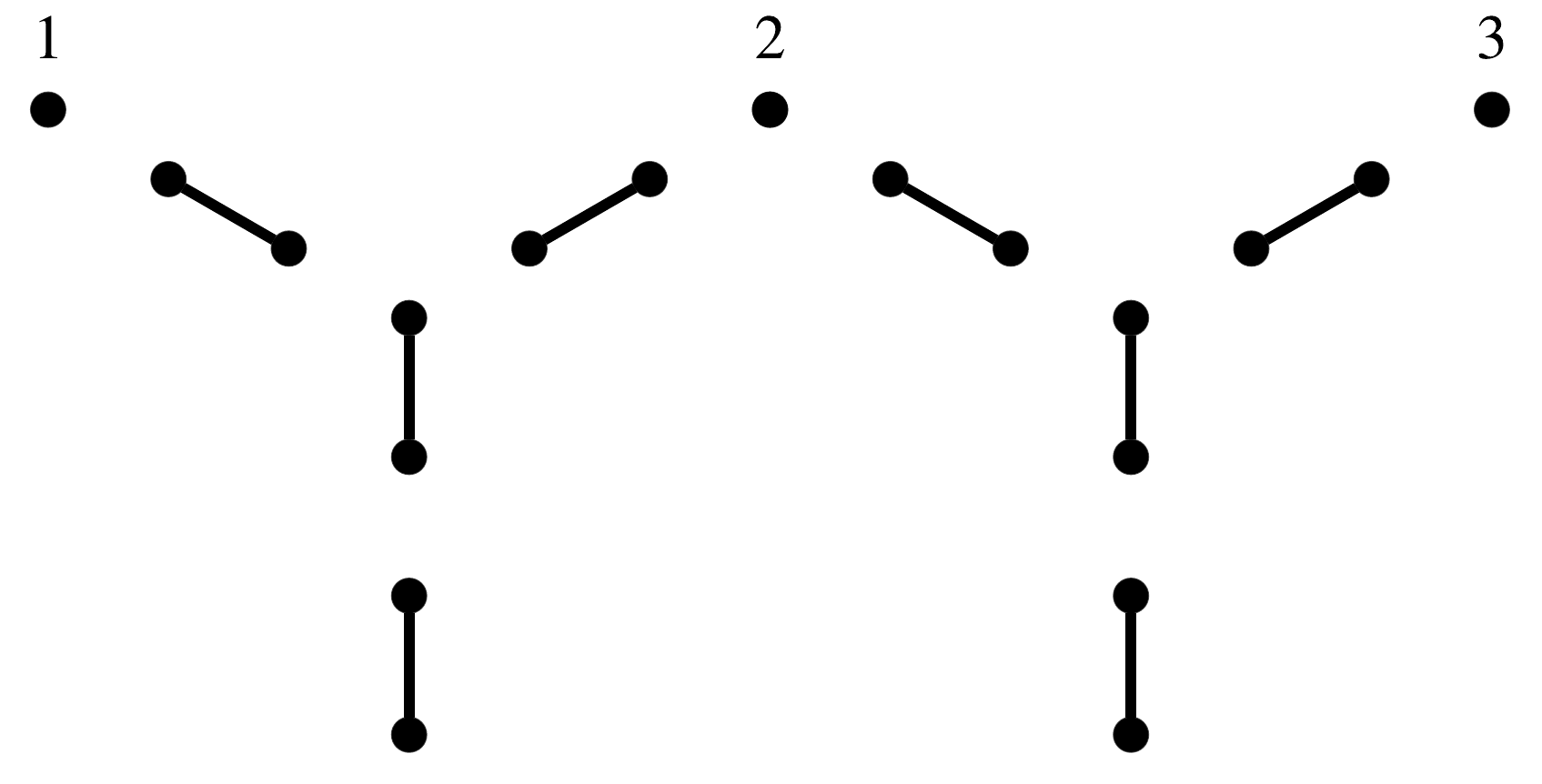}
\caption{An SSH double Y junction allowing for
braiding-based non-Abelian operations.
Three defects (1, 2, 3) define a three-dimensional
zero-energy subspace. 
The operation obtained by exchanging 
defects 1 and 2 first and 2 and 3 second is different
from the operation obtained by doing these two steps
in the reversed order.
}
\label{fig:YYjunction}
\end{figure}

The braid group on two strands, $B_2$, is Abelian, 
hence the image of any of its representation must be an
Abelian group as well. 
On the one hand, this is reassuring since $G_Y$ is Abelian. 
On the other hand, it also implies that $\pm \sigma_x$ and
$\pm \sigma_z$ can not be generated by braiding two defects, 
since their inclusion in the image of the representation would
make the image non-Abelian. 

Nevertheless, it is straightforward to find generalizatons of our setup, 
where more defects are present, and the quantum dynamics
induced by the various braidings 
of the defects generates a group of matrices
that is non-Abelian. 
An example with three defects is shown in Fig.~\ref{fig:YYjunction}; here
the worldlines of the defects are 
described by the braid group on three strands, $B_3$. 
The adiabatic 
counterclockwise exchange of the defects 1 and 2 results in the
overlap matrix 
\bean
Y_{12} = \left(\bna{ccc}
0 & -1 & 0 \\
1 & 0  & 0 \\
0 & 0 & 1
\eda \right), 
\eean
whereas the counterclockwise exchange of defects 2 and 3 
results in 
\bean
Y_{23} = \left(\bna{ccc}
1 & 0 & 0 \\
0 & 0  & -1 \\
0 & 1 & 0
\eda \right).
\eean
Clearly, the matrices $Y_{12}$ and $Y_{23}$ do not commute.

%=========
\section{Discussion}
\label{sec:discussion}
%=========

\emph{Experimental realizations.}
The topologically protected $Y$ gate
proposed above is presented in the context of a 
non-interacting non-superconducting tight-binding model. 
Because of the simplicity of the model, we expect that
the braiding dynamics simulated
here can be realized in experiments. 
Two suitable experimental platforms are 
 tunnel-coupled optical 
 waveguides\cite{SzameitReview,Martin},
and cold atomic systems where quantum states form a highly tunable
tight-binding lattice in momentum space\cite{MeierSSH,MeierAnderson}.
The role of hopping disorder have already been studied experimentally
in these setups\cite{Martin,MeierAnderson}.

\begin{figure*}
\includegraphics[width=2.0\columnwidth]{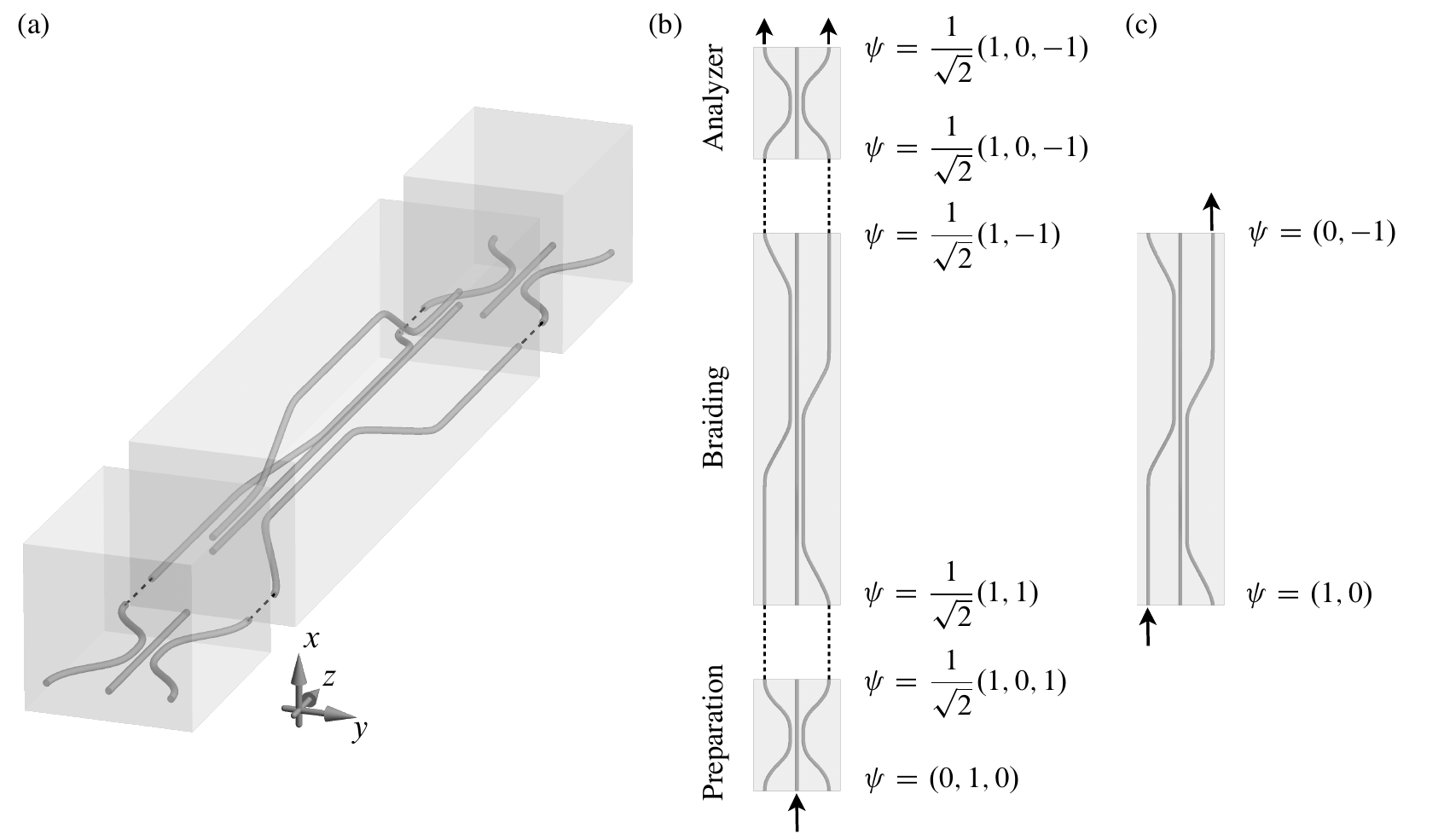}
\caption{
Optical waveguide structure for the 
demonstration of a braiding-based
$Y$ gate. 
(a) Three-dimensional view of the setup. 
The three segments (bottom: Preparation, middle: Braiding, top: Analyzer) 
are separated for better visualization, but 
they might have direct contact in a real device. 
(b) Details of the phase-sensitive measurement. 
Preparation segment creates a 50-50 superposition on 
the outer waveguides with no phase 
difference. 
Braiding segment acts as a $Y$ gate, shifting the relative phase 
between the two beams by $\pi$. 
Due to this phase shift and the corresponding complete destructive 
interference, the beam does not enter into the middle waveguide as passing 
through the Analyzer segment, but comes out with 50-50 intensity in the 
left and right outputs. 
(c) Braiding segment. Beam injected in left waveguide comes out from 
right output as a result of braiding. This measurement alone can not 
distinguish between the $X$ gate and a $Y$ gate.
Vector $\psi$ describes the 
input and output beam amplitudes at each segment.
}
\label{fig:OpticalWaveGuideMeasure}
\end{figure*}

Here, we discuss the optical waveguide array
where our model can be realized, 
and the effect of the braiding operation could be observed.
The setup, shown in Fig.~\ref{fig:OpticalWaveGuideMeasure}a,
 is inspired by a similar one shown in Fig.~2a
of Ref.~\onlinecite{KremerNonAbelian}.
This is a realization of our tight-binding
model with 4 sites, that is, with $N_c = 1$.

The waveguide array  
shown in Fig.~\ref{fig:OpticalWaveGuideMeasure} can be used to 
confirm that our braiding protocol 
provides a $Y$ gate. 
The experiment consists of two stages.
In the first stage, only the Braiding waveguide 
array is used (see Fig.~\ref{fig:OpticalWaveGuideMeasure}c).
This is an array of four waveguides, representing
the four sites in our tight-binding model 
with $N_c = 1$. 
The spatial distance between the waveguides is modulated
along the longitudinal, $z$ direction, mimicking the 
time-dependent modulation of the tunnel coupling 
in our tight-binding model\cite{KremerNonAbelian}.
Therefore, if a laser beam is injected in 
one of the two input waveguides, say, the left
one (bottom arrow of Fig.~\ref{fig:OpticalWaveGuideMeasure}c), 
then  the laser beam will come out 
from the right output waveguide 
(top arrow of Fig.~\ref{fig:OpticalWaveGuideMeasure}c)
as the result of braiding. 
Of course, this measurement is not revealing the $\pi$
phase shift of the output beam, so it is not able to 
distinguish between the $X$ gate and a $Y$ gate, 
for example. 

To make a phase-sensitive measurement, that is, to  
distinguish between an $X$ gate and $Y$ gate scenarios, 
one should do 
a second stage of the experiment, where the 
Braiding segment is supplemented by a
Preparation and an Analyzer segment of waveguides (see Fig.~\ref{fig:OpticalWaveGuideMeasure}b). 
These two segments are identical, but they serve different purposes. 
The structures are left-right symmetric, and serve as
perfect beam splitters with no phase difference
between the two output beams.
For example, the Preparation stage is used to convert an
incident laser beam entering the middle waveguide
(bottom arrow in Fig.~\ref{fig:OpticalWaveGuideMeasure}b)
to a 50-50 superposition output on the left and right waveguides,
with no phase difference. 
Then, this superposed laser beam will go through 
the Braiding segment performing a $Y$ gate, which 
will result in a similar balanced superposition
on the output ports, but will introduce a $\pi$ relative phase
shift between the split beams. 
Finally, the output of the Braiding segment is fed into the 
Analyzer segment, and due to the destructive interference 
arising from the $\pi$ phase difference, the beam will 
not penetrate the middle waveguide of the Analyzer but will
come out with 50-50 intensity in  its left and right output ports. 

In contrast,
if the Braiding segment would produce an $X$ gate, then
the output beam coming from the 
analyzer would come from the middle output port due
to constructive interference. 
Hence, observing a 50-50 intensity in the left and right output ports
of the setup in Fig.~\ref{fig:OpticalWaveGuideMeasure} is
a fingerprint of $Y$ gate. 
Note this two-stage experiment could be refined
to allow for a complete process tomography.
Also, based on earlier experiments, 
extending this setup to multiple waveguides in the 
presence of hopping disorder seems 
feasible\cite{Martin,menssen2019photonic}.

%Of course, any experimental setup has imperfections.
%Consider the example of realizing the braiding scheme in 
%the fully dimerized configuration, as described in section \ref{sec:dimerized},
%using a lattice of tunnel-coupled optical waveguides.
%In the experimental setup, the  waveguides representing the sites
%of the Y-junction will not be perfectly identical, hence, e.g., 
%the on-site energy parameter related to the geometry of these
%waveguides will fluctuate from site to site. 

%\emph{Absence of minigap.}
%In the Y-junction setup we study here, the number of sites
%on the A and B sublattices is different, 
%$N_A = N_B+ 2$, which guarantees the existence
%of the two-dimensional zero-energy subspace in the
%presence of chiral symmetry. 
%When chiral symmetry is broken in this Y-junction, 
%then a finite \emph{minigap} opens, which leads to 
%dynamical phase evolution in the quasi-zero-energy 
%two-dimensional subspace,
%as we have shown in 
%Fig.~\ref{fig:DistanceWithHoppingAndOnsiteDisorder}b.
%This implies that the final state as a function of braiding time $T$
%does not converge to a specific state in the $T \to \infty$ limit. 
%We emphasize that this conclusion can also be relevant
%for braiding schemes where the chiral symmetry is preserved, but
%the cardinality of the two sublattices are the same, $|A| = |B|$. 
%The role of this dynamical phase due to the small hybridization
%of zero modes was also discussed in the context of 
%the Majorana Y-junction\cite{Sekania}.

\emph{Numerical calculations}.
In all our results presented above, 
the overlap matrix was calculated by discretizing  the time
evolution operator, assuming a time-independent 
Hamiltonian for each time step. 
All numerical results were obtained by 
splitting the braiding time $T$ to 1000 time steps of
equal duration.
We performed a convergence test by halving the time step 
(doubling the number of time steps).
We have compared the error vs. braiding time curves for
the ordered cases as well as a few realizations of disorder, 
without observing any significant difference between 
the curves obtained with 1000 time steps and those with 2000 time steps.

%=========
\section{Conclusion}
\label{sec:conclusion}
%=========

In conclusion, we have proposed a single-particle
system, the SSH Y-junction, where braiding of defects
and the corresponding zero modes
provides topologically protected quantum gates.
The topological protection of the gates arises as the joint consequence
of the chiral and the time-reversal symmetries
of the Hamiltonian, and the
spatial separation of the zero modes. 
Cyclic adiabatic deformations of the Hamiltonian establish
a group representation between the braid group 
and the available group of quantum gates, the latter being
Abelian when there are only two defects in the system, 
but non-Abelian if at least three defects are present. 
Even though the model system introduced here will probably not
be used in practical topological quantum computing schemes, 
it does stand out from earlier schemes
because of its conceptual simplicity, 
and its strong potential for experimental realization.

%\emph{Note:} During the completion of this work, we became
%aware of a related manuscript, where the minimal version of the
%SSH Y-junction ($N_c = 1$) was introduced.\cite{Droth} 
%That manuscript contains significant contributions from 
%two of us (P.~B. and A.~P.), but it was prepared and submitted without
%our knowledge and without our consent.

\acknowledgments
We thank M.~Droth, M.~Kremer, L.~Teuber, A.~Szameit, D.~Varjas for useful discussions,
and G.~Tak\'acs for computational resources. 
This research was supported by the National Research Development and
Innovation Office of Hungary within the Quantum Technology National
Excellence Program (Project No. 2017-1.2.1-NKP-2017-00001), 
and Grants FK124723, K115575, and K115608.
This work was completed in the ELTE Institutional Excellence Program 
(1783-3/2018/FEKUTSRAT) supported
by Hungarian Ministry of Human Capacities.
O.~L.~was supported by the
Bolyai and Bolyai+ scholarships of the Hungarian
Academy of Sciences. 

\appendix

\section{Zero-energy eigenvectors of
a chiral symmetric Hamiltonian with sublattice imbalance}
\label{app:darkstatetheorem}

In the main text, we described simple single-particle
Hamiltonians with chiral symmetry, where the 
two sublattices contained a different number of sites, 
$N_A = N_B + 2$. 
We claim that such Hamiltonians have 
a twofold degenerate zero-energy subspace. 
Here we provide a simple proof. 

\renewcommand{\vec}[1]{\text{\boldmath{$ #1 $}}}
\newcommand\at[2]{\left.#1\right|_{#2}}
\newcommand{\matr}[1]{\begin{pmatrix} #1 \end{pmatrix}}

Take a Hamiltonian of
size dimension $N_A + N_B$,
with the special form
\begin{equation}
	H = \matr{0 & V\\ V^\dag & 0},
\end{equation}
where $V$ is a $N_A\times N_B$ matrix 
with complex entries, 
couples the two sublattices, $A$ and $B$.
Without the loss of generality, assume $N_A > N_B$.
For example, in the model shown in Fig.~\ref{fig:DimerizedYJunction}a,
we have $N_A = 6$ and $N_B = 4$.
Any such matrix $V$ 
can be factorized as
$V= L W R^\dag$, where 
$L$ is a $N_A\times N_A$ unitary matrix, 
$R$ is a $N_B\times N_B$ unitary matrix, and 
$W$ is a $N_A\times N_B$ rectangular diagonal matrix,
meaning that all entries but the ones of the form $W_{i,i}$ 
are zero;
this is known as \emph{singular value decomposition}. 
Using this decomposition, it is easy to see that
\begin{equation}
	H = \underbrace{\matr{L&0\\0&R}}_{U} \underbrace{\matr{0 & W \\ W^\dag & 0}}_{\tilde{H}}  \underbrace{\matr{L^\dag&0\\0& R^\dag}}_{U^\dag},
\end{equation}
where $U$ is an unitary matrix (not to be confused with
the propagator in the main text).

The spectrum of $H$ and $\tilde H$ is the same, since they 
are related by the unitary transformation $U$. 
From the structure of the rectangular block $W$ discussed above, 
it is clear that $\tilde H$ has $N_A - N_B$ 
zero eigenvalues, and hence that is also true for 
$H$. 
That concludes the proof. 

Note that this claim and its generalizations are sometimes
called \emph{dark-state theorem}, and 
the eigenstates of $H$ belonging to the zero-energy
subspaces are called \emph{dark states}. 
See also Ref.~\onlinecite{Morris} for a related discussion. 
Furthermore, these considerations can be straightforwardly 
generalized to the case when the two diagonal blocks 
of $H$ are not zero but proportional to the identity matrix,
with potentially different proportionality constants.

%===

\bibliography{SSHBraidingManuscript}

\end{document}